\documentclass[preprint,authoryear,12pt]{elsarticle}

\usepackage{amssymb}

\journal{Nuclear Physics B}

\begin{document}

\begin{frontmatter}
\title{Formation of dwarf ellipticals and dwarf irregular galaxies by interaction of giant galaxies under environmental influence}


\author[1]{Tanuka Chattopadhyay}
\ead{tanuka@iucaa.ernet.in}
\author[1]{Suma Debsarma}
\ead{debsarma@rediffmail.com}
\address[1]{Department of Applied Mathematics, Calcutta University, 92 A.P.C Road, Calcutta 700009}
\author[1]{Pradip Karmakar}
\author[2]{Emmanuel Davoust}
\address[2]{Institut de Recherche en Astrophysique et
Plan\'etologie, Universit\'e de Toulouse/CNRS, 14 Avenue Edouard
Belin, 31400 Toulouse, France } \ead{edavoust@irap.omp.eu}
\begin{abstract}

A model is proposed for the formation of gas-rich dwarf irregular

galaxies and gas-poor, rotating dwarf elliptical galaxies

following the interaction between two giant galaxies as a function

of space density. The formation of dwarf galaxies is considered

to depend on a random variable, the tidal index $\Theta$, an
environmental

parameter defined by Karachentsev et al. (2004), such that for

$\Theta < 0$ the formation of dwarf irregular galaxy is assured

whereas for $\Theta > 0$ the formation of dwarf ellipticals is

preferred. It is found that for particular ranges of the

interactive parameters the model predictions are in good agreement

with the observed number density of the different galaxy types as

a function of space density in four clusters of galaxies. This

supports the fact that galaxy interactions do not all necessarily

give rise to the formation of either dwarf irregulars or dwarf

ellipticals. It is also shown that the formation of dwarf

irregulars at high densities is  much lower than that of dwarf

ellipticals, and that the formation of the latter reaches a

maximum at a particular space density, unlike the former. This

suggests that at high densities many dwarf irregulars are stripped

of their gaseous envelopes to become dwarf elliptical.

\end{abstract}


\end{frontmatter}


\section{Introduction}

Dwarf galaxies are small, low-luminosity and low-metallicity

galaxies. In spite of their vast majority over other morphological

types (Sandage \& Binggeli 1984; Ferguson \& Binggeli 1994; Mateo

1998) their formation scenario is still far from understood. Giant

galaxies generally form either through the gravitational collapse

of a huge protogalactic gas cloud (Larson 1974; Dekel \& Silk

1986; White \& Frenk 1991; Frenk et al. 1996; Kauffmann et al.

1997), by merger of two or more disc galaxies (Toomre 1977;

Ashmann \& Zepf 1992; Zepf et al. 2000), by multiphase dissipative

collapse (Forbes et al. 1997), by dissipationless merger model, or

by accretion and in situ hierarchical merging (C\^ot\'e et al.

1998; Mondal et al. 2008; Chattopadhyay et al. 2009).  Dwarf

galaxies, on the other hand, occupy a separate region in the

Fundamental Plane (Kormendy 1985; Chattopadhyay et al. 2012;

Chattopadhyay \& Karmakar 2013), which suggests that they

have a different formation mechanism.\\

Some dwarf galaxies might be formed by galaxy collisions because

dwarf irregular galaxies (hereafter dIrrs) have been found in the

tidal tail of interacting galaxies (Schweizer 1982; Bergvall \&

Johansson 1985; Mirabel et al. 1991; Braine et al. 2000;

Weilbacher et al. 2000; Ferreiro et al. 2005; Mendes de  Oliverira

et al. 2000; Mundell et al. 2004; Sheen et al. 2009;  Kaviraj et

al. 2012). The formation of tidal dwarf galaxies has also been

demonstrated by numerical simulations (Barnes \& Hernquist 1992;

Bournaud \& Duc 2006; Dabringhausen \& Kroupa 2013). The formation

of tidal dwarf galaxies from stripped gas has been reviewed by

Bournaud (2010). Kroupa (2012) has shown that the existence of two

types of dwarf galaxies (with or without dark matter) is

incompatible with the current standard model of cosmology. The

formation of ultra-compact dwarf galaxies (which have masses in

the range $10^6 M_{\odot}$ to $10^8 M_{\odot}$ and radii 10 pc

-100 pc) is a different issue (Dabringhausen et al.

2008;Chattopadhyay et al. 2012;Chattopadhyay \& Karmakar 2013). \\

The formation of dwarf ellipticals (hereafter dEs), on the other

hand, is probably related to early-type galaxies (E or S0),

because the frequencies of dEs and giant early types are
increasing

functions of space density (Binggeli et al. 1990; Ferguson \&

Binggeli 1994). Also van Zee et al. (2004) have observed 16 dEs in

the Virgo cluster and found that they are similar to dIrrs in

rotation amplitude and luminosity. These observations suggest that

some cluster dEs may be formed when the neutral gaseous medium of

dIrrs is stripped under the influence of some environmental factor

(Dunn 2010). Investigating the formation of dwarf galaxies as a

result of giant galaxy interaction and the condition under

which dIrrs and dEs are formed is thus now a priority. \\

In the present paper we develop a numerical statistical simulation

model for the formation of dIrr and dE galaxies along with giant

galaxies as a function of increasing space density. This is a more

extended version of the galaxy interaction models of Silk \&

Norman (1981) and Okazaki \& Taniguchi (2000). We have considered

a random production of dIrr and dE galaxies during each
interaction on

the basis of an environmental parameter, the tidal index $\Theta$

(Karachentsev et al. 2004). We describe the model in section 2,

the observations are presented in section 3, and the results and

interpretations are given in section 4.

\section{Model}

The present model is based on the scenario of fragmentation

and hierarchical clustering scheme of galaxy formation. According

to this scenario, the primordial density fluctuation at

recombination is  $ \delta\rho / \rho \sim (M/M_0)^{-\alpha}$
(Efstathiou 1979),

where $\alpha \sim 1/2$ for (for $\Omega \sim 1$) or $\alpha \sim
1/3$ (for $\Omega \sim 0.1$).

Hence $M_o \sim 10^{14.9 - 3/\alpha} \Omega^{1 - 1/\alpha} h^{-1}$
(h = $H_o/50 km s^{-1} Mpc^{-1}$).

So a primordial spectrum of isothermal density fluctuation in the

early universe could form bound clouds of mass $M_0$ at
z$\sim$1000

i.e. $M_0 \sim 10^8 M_{\odot} (\Omega \sim 0.1)$ to $10^9
M_{\odot}(\Omega \sim 1)$. Such clouds can survive

for times longer compared to their collapse time scales(SN81).

These gas clouds are clustered gravitationally forming gas rich

protogalaxies at a late epoch (z $\le $ 10) (Peebles 1974 ;

Davis \& peebles 1977; Gott et al. 1979).The gas clouds that

form protogalaxies interact dissipatively and dissipation leads to

the formation of the luminous parts of galaxies. The dark material

or the parent material of these clouds play an important role

in forming galaxy clusters and haloes. White and Rees (1978)

have shown that the growth of density fluctuations on larger

scales can be described by a spectrum $\delta \rho /\rho =
(M/M_i)^{1/2 - n/6} (t/t_i)^{2/3}$

which leads to $M/M_i \sim (R/R_i)^{6/(n+5)}$ in an
Einstein-de-Sitter Universe

($\Omega \sim 1$). This leads to $M_i \sim 5 \times 10^{14}
(1+z_i)^{-2} M_{\odot}$ and $R_i \sim R_0 (1+z_i)^{-5/3}$

for an observed galaxy correlation function and n=0. $z_i (\le
10)$

is the epoch of galaxy clustering. The protogalaxies ultimately

form clusters of galaxies at $z\le 10$. Hence the onset of galaxy
clustering

is generally considered at $z \sim 5 - 10$. During the initial

phase of galaxy clustering many collisions among these
protogalaxies

will occur and those result in mergers.\\

Dressler(1980) has observed the fraction of galaxies of various

morphological types as a function of local density in rich
clusters.

For instance in small groups and other low density regions, the
fraction

of bright galaxies which are spiral, is around 80 \%, with about
20\%

S0s and very few ellipticals. Moving inward from the outskirts of
a cluster,

the fraction of spirals decreases steadily while the S0 fraction
rises

steadily but is rapidly caught up by the elliptical fraction once

high densities are reached. \\

Hence following the study of Dressler(1980)

we have assumed that initially at the onset of the galaxy
clustering

epoch (lowest space density) all the galaxies are protospirals.

A model of protogalaxy interaction was first introduced by

Silk \& Norman(1981) and it reproduced the observed fraction of

galaxies of various morphological types as a function of logarithm

of the ratio of S0s to spiral. This ratio is proportional to the

space density. Subsequently Okazaki \& Taniguchi (2001) have also

considered merger of such protogalaxies and in addition each

interaction in their model was accompanied with a production of

dwarf galaxy whose morphology was not specified. All the previous

models considered that initially at the epoch of galaxy clustering

all the protogalaxies were protospirals. \\

The present model, like the one of Okazaki \& Taniguchi (2001),

considers the production of a dwarf galaxy as

a result of protogalaxy interaction, but each time it is produced

its morphology is determined on the basis of an environmental
parameter

$\Theta$ which is a random variable defined by Karachentsev \&
Makarov

(1988). $\Theta$ describes the local mass density around a galaxy
$i$

as $\Theta_i = max [ log (M_k /D_{ik})^3 ] + C, i = 1,2,....N$

where $M_k$ is the total mass of any neighbouring galaxy

separated from the considered galaxy by distance $D_{ik}$. In a

previous paper (Chattopadhyay et al. 2010) we classified dwarf

galaxies on the basis of several parameters and found two groups,

one dominated by dIrrs and the other by dEs, and for these two

groups the average value of $\Theta$ was negative and positive,

respectively. The observed values of $\Theta$, taken from

Karachentsev et al.(2004), are given in Table 1 along with galaxy

names and morphological indices T. \\

The distribution of $\Theta$ has been successfully fitted by a

Gamma distribution (Anderson Darling value, which is a measure of

goodness of fit, is 0.65). Since the Gamma distribution is valid

for positive values only we have used a coordinate transformation

Y = X+3 to make all observed $y_i\ge 0$. Here X and Y are two

random variables. In particular X stands for $\Theta$. $y_i$ ($i$
=

1,2,....) are the observed values of Y found by the above

translation over all observed values of $\Theta$. Then after

fitting the distribution we used the back substitution X = Y - 3

for getting the original values of the variables. Now for each

iteration we have generated $\Theta $ randomly (See Appendix 1)

from the Gamma distribution (Fig. 1). If $\Theta <$ 0 we assume

that a dIrr galaxy has been produced and if $\Theta >$ 0 , a dE

has been produced.\\

Hence the present model includes the formation of both dIrrs and

dEs as a result of tidal interactions between giant galaxies,

either Spirals, Lenticulars or both. This is warranted by numerous

observational studies of dwarf galaxies found in interacting or

merging giant galaxies (Bergvall \& Johansson 1985; Mirabel et al.

1991; Hibbard \& Higdon 2001;  Delgado-Donate et al. 2003;

Knierman et al. 2003; Mundell et al. 2004;  Allam et al. 2007; Duc

et al 2007; Kaviraj et al. 2012). We consider the production of

dEs and dIrrs separately, rather than that of dwarf galaxies in

general. Both types of dwarf galaxies can be produced in each

interaction and the production of either type is random. Our model

also includes the production of ellipticals and/or of lenticulars.

The case of forming an elliptical in the merger of two spiral

galaxies (Ashman \& Zepf 1992; Zepf et al 2000) had not been

considered in the previous papers on this subject. The scheme of

interaction assumes that at each interaction a dwarf galaxy is

produced, and its type depends on $\Theta$ which is drawn
randomly.

This scheme is summarized in Table 2.  For a draw of any

random value of $\Theta$, if $\Theta > 0$ we assume that $k_1$ dEs
are

produced and if $\Theta < 0$ $ k_1^{'}$ dIrrs are produced.

Since the production of different types of dwarfs is random

we have used a 'slash' in each interaction of the merger tree.

In Table 2, the parameters a, b, c, d, and e are the probabilities
with

which the impact between protogalaxies occur and are called impact

parameters. Based on the scheme of Table 2, the kinetic equations

for the evolution of each morphological type reduces to,\\

\begin{equation}
 \frac{1}{\gamma} \frac{d n_{Sp}}{dt} = -2 a \ n_{Sp}^2 - n_{S0} n_{Sp}
\end{equation}

\begin{equation}
 \frac{1}{\gamma} \frac{d n_{S0}}{dt} =  a \  n_{Sp}^2 - b \ n_{S0} n_{Sp}- d \ n_{S0}^2 + e \  n_{S0} n_{Sp}
\end{equation}

\begin{equation}
 \frac{1}{\gamma} \frac{d n_{E}}{dt} =  (1-a) \  n_{Sp}^2 + b \ n_{S0} n_{Sp}+ c \ n_{S0} n_{Sp} + d \ n_{S0}^2
\end{equation}

\begin{equation}
 \frac{1}{\gamma} \frac{d n_{dIrr}}{dt} =  [k_1^{'} \ a + k_2^{'} \ (1-a)] \  n_{Sp}^2 + [k_3^{'} \ b + k_4^{'} \ c + k_5^{'} \{ 1 - (b + c + e) \}] \ n_{S0} n_{Sp} + [d \ k_6^{'} + (1-d) \ k_7^{'}] \ n_{S0}^2
\end{equation}

\begin{equation}
 \frac{1}{\gamma} \frac{d n_{dE}}{dt} =  [k_1 \ a + k_2 \ (1-a)] \  n_{Sp}^2 + [k_3 \ b + k_4 \ c + k_5 \{ 1 - (b + c + e) \}]
 \ n_{S0} n_{Sp} + [d \ k_6 + (1-d) \ k_7] \ n_{S0}^2
\end{equation}

\noindent where $n_{Sp}, n_{S0}, n_E, n_{dIrr}, n_{dE}$ are the

number densities of Spirals (Sps), lenticulars (S0s), ellipticals

(Es), dIrrs and dEs respectively, $\gamma$ is the mean collision

rate and $k_i$ (i = 1-7) and $k_i^{'}$ (i = 1 - 7) are the numbers

of respective dEs and dIrrs formed in each collision. The

differential equations were constructed from the merger tree. For

example, equation (2) gives the change in the number density of

S0s with time. S0s decrease in number with probability b when one

Sp and one S0 interact and with probability d when two S0 galaxies

interact. The number density is increased with probability a when

two Sps interact, and with probability e when one S0 and one Sp

interact. Similarly the remaining equations follow. Concerning

$dn_{Sp}/dt$ we have assumed that the decrease in spiral galaxies

is dominated by the production of S0s more than by that of Es

(Dressler et al. 1997; Ellis et al. 1997; Ghigna et al. 1998; van

Dokkum 2002). In order to solve equations (1) to (5), we introduce

the variable $ x = n_{Sp} /n_{S0}$ which decreases monotonically

with increasing galaxy density (Silk \& Norman 1981 ; Okazaki \&

Taniguchi 2000). Then we obtain the implicit solution as,\\

$$ \frac{n_{Sp}}{n_0}=[ \{\frac{ax^2}{ax^2+(2a+e-b)x+(1-d)}
 \}^{\frac{1}{2(1-d)}}]$$
\begin{equation}
.exp[\frac{(2a+e-b)d}{(1-d)a
\Delta}(tan^{-1}\{\frac{2ax+(2a+e-b)}{2a\Delta} \} -
\frac{\pi}{2})]
\end{equation}

\begin{equation}
n_{S0} = \frac{n_{Sp}}{x}
\end{equation}

\begin{equation}
n_E = \int_x^{\infty}\frac{n_{Sp} \{ (1-a)x^2 + (b+c)x + d \}}{x^2
\{ ax^2 +(2a+e-b)x + (1-d) \} } dx
\end{equation}

\begin{equation}
n_{dE} = \int_x^{\infty} \frac{(Ax^2 +Bx + C) n_{Sp} }{x^2 \{ ax^2
+(2a + e - b)x + (1-d) \} } dx
\end{equation}

\begin{equation}
n_{dIrr} = \int_x^{\infty} \frac{(A^{'}x^2 +B^{'}x + C^{'}) n_{Sp}
}{x^2 \{ ax^2 +(2a + e - b)x + (1-d) \} } dx
\end{equation}

\noindent where $ A = k_1a + k_2(1-a)$, \

$B = k_3b + k_4c + k_5 \{ 1 - (b+c+e) \}$ and

$ C = d k_6 + (1-d) k_7 $\\

\noindent and $ A^{'}, \ B^{'}, C^{'}$ are identical with $A$, $B$
and $C$ respectively, except for replacing $k_i$ (i= 1-7) by
$k_i^{'}$ (i= 1-7).

$ \Delta^2 = \frac{(1-d)}{a} - \{ \frac{(2a +e -b)}{2a} \}^2 $.

Since $\Delta^2 > $ 0, $ \frac{(b-e)}{2} < a < 1$  and also

we have $ 0 \le a \le 1, \  0 \le b+c+e \le 1, \  0 \le d \le 1$.

(see the Appendix 2 for the actual derivation of equations (6) to (10)).\\

We have mentioned that initially all galaxies are assumed to be

spiral galaxies and their initial number density is $n_0$.

Initially we assume $ k_i = k_i^{'}$ (i = 1-7) but then for each

increment $\Delta x$ of $x$ we take decrements of $ k_i$ and

$k_i^{'}$ as $\Delta k_i$ and $\Delta k_i^{'}$, if corresponding

to $\Theta$ a dE or dIrr galaxy is formed, otherwise there is no

decrement in $k_i$ or $k_i^{'}$ (i =1-7) respectively. Our

objective is to find plausible ranges of the impact parameters

and values of $k_i$ and $k_i^{'}$ (i= 1-7) for which the computed

number densities of galaxies of different morphological types

as a function of log(1/x) are consistent with the number densities

of the observed ones i.e. the $\chi^2$ goodness of fit value

between observed and simulated values are small enough

($\chi^2 < 1 $ with high p-value).

\section{Data}

We have used three data sets, which are described in Table 3. The

first one consists of one sample in the Virgo cluster and four in

Coma (Michard \& Andreon 2008). Coma0 is the full sample of Coma

galaxies, whereas Coma1, 2 and 3 consist of galaxies at increasing

distances from the cluster center. The second one is a sample in

the Centaurus cluster (Jergen \& Dressler 1997). In this sample,

irregular galaxies having magnitudes $M_B \ge -10$ are considered

as dIrrs. The last sample is that of 2096 galaxies in the Virgo

cluster (Binggeli et al. 1985). In this sample, irregular galaxies

with magnitudes  $M_B \ge $ -16 are considered as dIrrs. In Table

3, the parameter $x$ is the ratio of spirals to lenticulars
defined above.

\section{Results and discussions}

In order to find the best model we minimize the sum of squares

of the deviations between predicted and observed number of

galaxies of various morphological types weighted by the predicted

number of galaxies i.e. $\Sigma_{i=1}^5 \Sigma_{j=1}^N
\frac{(n_{pred} - n_{obs})^2_{i,j}}{n_{pred}} = \chi^2$

where $i$ stands for the five morphological types (viz. Sp, S0, E,
dE, dIrr)

and $j$ stands for the data points of groups and clusters given

in Table 3 , $N$ is the total number of galaxies of each type.

$n_{pred}$ (found from equations (6)-(10)) is the number of

predicted galaxies and $n_{obs}$ (found from Table 3) is the

number of observed galaxies. Table 4 shows the best ranges

of impact parameters a, b, c, d, e and values of $k_i$ and

$k_i^{'}$ (i=1-7) for which $\chi^2$ is small enough together

with a high level of significance (characterized by p-value, p=
0.995).\\

Our simulations show that the fraction of spiral galaxies

decreases very slowly up to $log(1/x)$ = -0.5 and after

that point the fall is almost exponential and the $\chi^2$

goodness of fit value with an exponential curve gives a very

small value of $\chi^2 \sim$ 0.12 for -0.5 $< log(1/x)\le 2$.

We remind the reader that $log(1/x)$ increases monotonically

with increasing space density. The fractions of E and S0 galaxies

have moderate maxima around log(1/x) = -0.5 and 0 respectively.

The maximum of $log(1/x)$ for Es occurs at a lower space density

than that of S0s. This must be due to the formation process of the

latter, by gas stripping in Spirals, which is favored by high

density environments. The peak for S0 galaxies at $log(1/x)$ = 0
is

close to the value found by Okazaki \& Taniguchi (2000), but the
peak

for ellipticals is at a lower value than the one found by them.\\

In our simulations, the fraction of dE galaxies has a pronounced
peak

at $log(1/x)\simeq 0$  and it decreases at higher values of
log(1/x),

i.e. as log(1/x) tends to 2 the fraction is close to zero

(Fig.2, green solid line).  The fraction of dIrrs is much smaller

than that of dEs and it is

also a decreasing function of space density.  The low number of
dIrrs

compared to dEs at medium densities may be due to the fact that

gas stripping from a large number of dIrrs by neighboring galaxies

leads to the formation of dEs (van Zee et al. 2004). The fall in

the fraction of dEs at still higher densities may be due to the

decreasing fraction of spiral galaxies in the high-density regime

taking part in the interaction.  There are also very few dwarfs in
regions

of low space density. This is consistent with observations

(Yahil et al.1998; Shanks et al 2001; Mathews et al. 2004). \\

The observed fraction of galaxies

of various morphological types is in the range $-0.5 <log(1/x) <
0.5$

(see Fig.2 and Table 3).  Ann (2007) observed various dwarf

galaxies in the local Universe and derived the local density

($\rho$) by calculating the mass of galaxies with a projected

distance of 1 Mpc and a line of sight velocity difference

less than 500 km/s assuming constant M/L.  Ann (2007) plotted the

fraction of 'dE,N', Im, dS0 and dE versus log($\rho$) and found
that

the majority of dE galaxies are located in the overdense regions,

with a peak around $log(\rho)\sim 1$. This is similar to our
result

that dE occurs in dense regions (Fig.2) i.e. in clusters of

galaxies rather than in the field. \\

We now turn to the number of galaxies of various morphological
types

produced in our simulations.  dEs and dIrrs have their maxima
around

$k_i$ = 2 and $k_i^{'}$ = 1, (i = 1-7) respectively from best
fitting.

This is almost consistent with the observations of dwarf galaxies
in

individual galaxy interaction (Table 5) i.e. the maximum values

of $k_i$ and $k_i^{'}$ (i =1-7) in Fig.3 are 2 and 1 which are

close to the values 1/2/3 for most of the observations in

Table 5. It is to be mentioned that the type of dwarf galaxies

has not been specified in Table 5.  In the highest density

regime (viz. $log(1/x)$ in the range 0 - 2), the formation rates

of dwarf galaxies of both types in individual interaction are

roughly constant (viz. $k_i \sim 2$ for dEs and $k_i ^{'} \sim 1$

for dIrrs). \\

The curve for $k_1$ (in magenta) runs higher than the other curves

in the top right graph of Fig.3, whereas the curve for $k_4^{'}$
(in cyan)

runs higher than the other ones in the bottom right graph of
Fig.3.

Now $k_1$ is the number of dEs formed from Sp+Sp interaction

(viz Table 2) and $k_4^{'}$ is the number of dIrrs formed from

Sp+S0 interaction. Hence the formation

of dIrrs is preferred in Sp+S0 interaction compared to Sp+Sp for

dEs. Kaviraj et al.(2012) have performed a statistical

observational study of tidal dwarf populations using a

homogeneous catalogue of galaxy mergers from SDSS. They found

that mergers producing tidal dwarfs involve two progenitor spirals

in 95\% of the cases and at least one progenitor spiral in the
remaining

cases. The fraction of tidal dwarfs where both parents are of

early type is less than 2\%. These results are almost consistent
with

the results in the present paper. \\

The main innovation of the present model, compared to those of

Silk \& Norman (1981) and Okazaki \& Taniguchi (2000) is that we

are able to trace separately the formation of dEs and dIrrs as a

function of increasing space density using the environmental

parameter $\Theta$. This is important because the formation of dE

or dIrr galaxies constrained by the tidal environment remained

controversial. It has been claimed that dEs are transformed from

dIrrs when enough gas is expelled as a result of tidal influence

by neighbouring galaxies (van Zee et al. 2004; Dunn 2010). Hence

the question remained how these two types of dwarf galaxies are

formed. We have incorporated the above idea in the present model

by assuming that the formation of dEs or dIrrs depends on a random

variable $\Theta$ (Karachentsev et al. 2004). We indeed found in
an earlier

paper (Chattopadhyay et al. 2010) that dE galaxy formation is
preferred

when $\Theta >0$ and that of dIrrs when $\Theta <0$.  We also
considered

values of $log(1/x)$ up to 2, thus reflecting the proper trend of

number density of the different galaxy types as a function of

increasing space density, whereas Okazaki \& Taniguchi (2000) only

considered values of $log(1/x)$ up to 1.

In conclusion, assuming that most of the present-day dwarf

galaxies originate from galaxy interactions, we have been

able to find model parameters (a, b, c, d, e) which give results

in agreement with the observations in different environments. This

suggests that most present-day dwarfs were formed in protogalaxy

interactions, and that they are not building blocks of the

hierarchical formation of galaxies.

\section{Acknowledgements}

T. C. thanks DST, India for supporting her a Major Research Grant.

The authors are very grateful for the various points suggested by

the referee for improving the work. The authors are also grateful

to A.K. Chattopadhyay for his fruitful suggestions.

\section{ Appendix 1: Generation of random sample from Gamma distribution}

The probability density function (pdf) of the Gamma

distribution of a random variable $X$ (here $\Theta$) is

$f(x; k, \theta) = \frac{x^{k-1} e^{-\frac{x}{\theta}}}{\theta^k
\Gamma^{k}}$, for $x>0, \theta>0$

where $k$ is the shape parameter  and 1/ $\theta$ is its rate

parameter. Now for a given pdf $f(x)$, the cumulative distribution

function (cdf),  $F(x) = P(0\le X\le x) = \int _0^x f(x) dx.$

where $F(x)$ is a random number $r$ and $0 < r , 1$. Then $x =
F^{-1}(r)$

generates a sample value $x$ of $X$ for a randomly chosen $r$.

Since the Gamma distribution is one of the many distributions from

which it is difficult or even impossible to directly simulate by
the

above inverse transform, we use the Accept-Reject method by which

samples can be generated from the target density $f(x)$ through

another density $g(x)$, known as the instrumental density. \\

If $f$ is the density of interest (which is Gamma in our case)

known as target density on an arbitrary space we can write

$ f(x) = \int _0^{f(x)} dr$. Thus $f$ appears as the marginal

density of the joint distribution $(x,R) \sim uniform \{(x,r),
0<r<f(x)\}$.

Hence simulating $X \sim f(x)$ is equivalent to simulating

$(X,R) \sim$ uniform$\{(x,r), 0<r<f(x)\}$. As the simulation

of the uniform pair ($X,R$) is often not straightforward

we use the following simplification  (Robert \& Casella 2004)

Let $ X \sim f(x)$ and let $g(x)$ be a density function

that satisfies $f(x) \le M g(x)$ for some constant$M\ge 1$.

Then to simulate $x \sim x$ it is sufficient to generate

$Y \sim g(y)$ and $R|Y \sim $uniform$(0, Mg(y))$ until $0<r<f(y)$

Thus we have the algorithm:

Step 1: Generate $X \sim g$ and $R \sim$ uniform$(0,1)$

Step 2: Accept $Y = X$ if $ r < \frac{f(x)}{M g(x)}$

Step 3: Return to Step 1 otherwise.

For our situation the target density $f(x)$ is $Gamma(k,\theta$).

We take the instrumental density $g(x)$ as $Gamma (a,b)$ where

$a=[k]$ and without loss of generality we assume $\theta =1$.

Then $ M = b^{-k} (\frac{k - a}{(1-b)e})^{k -a}$

The optimum choice of $b$ for simulating from $Gamma(k,1)$

is $ b = \frac{a}{\alpha}$.

\section{ Appendix 2: Derivation of equations 6 to 10}

\noindent Dividing equation (1) by equation (2),

$$\frac{dn_{Sp}}{dn_{S0}} = \frac{-2an_{Sp}^2 - n_{S0}
n_{Sp}}{a n_{Sp}^2 - b n_{S0} n_{Sp} - d n_{S0}^2 + e n_{S0}
n_{Sp}}$$

1.e. $$dn_{Sp} = \frac{-2ax^2 - x}{ax^2 + (e-b)x - d}
 (\frac{1}{x} dn_{Sp} - \frac{n_{S0}}{x} dx)$$

 [Since, $$ x = \frac{n_{Sp}}{n_{S0}} \ i.e. \ dn_{Sp} = x dn_{S0} +n_{S0} dx \ i.e. \ dn_{S0} = \frac{1}{x} dn_{Sp} - \frac{n_{S0}}{x}dx ]$$

 i.e. $$ \{ 1 + \frac{2ax + 1}{ax^2 + (e-b)x - d } \}dn_{Sp} = \{ \frac {
 2ax + 1}{ax^2 + (e-b)x - d} \} dx \ (\frac{n_{Sp}}{x})$$

 i.e.$$ \frac{dn_{Sp}}{n_{Sp}} = \frac{2ax +1}{ \{ ax^2 + (2a + e -b) x + (1 - d) \}x}$$

 Integrating between $\infty$ and x and assuming at x = $\infty$ ,
 $n_{Sp} = n_0$,

 $$ ln (\frac{n_{Sp}}{n_0}) = - \int_x^{\infty} \frac{(2ax + 1)
 dx}{\{ ax^2 + (2a + e - b)x + (1-d)\}x}$$

 i.e. $$ln (\frac{n_{Sp}}{n_0}) = - \int_x^{\infty} [ \frac{1}{x}
 - \frac{1}{2} \{ \frac{2ax + (2a + e -b) + (e - b -2a)}{ax^2 +(2a
 + e - b)x + (1-d)}\}$$
 $$ + \frac{d}{(1-d)}\{ \frac{1}{x} - \frac{ax + (2a + e - b)}{ax^2 +(2a + e - b)x + (1-d)}\}]dx$$

 i.e.$$ln (\frac{n_{Sp}}{n_0}) = - [ \frac{1}{2} \ ln \{
 \frac{x^2}{ax^2 + (2a + e -b)x + (1-d)}\}$$
 $$ + \frac{1}{2} (\frac{d}{1-d}) \ ln \frac{x^2}{ax^2 + (2a + e
 -b)x +(1-d)}]_x^{\infty}$$
 $$ - \frac{(2a + e - b)d}{(1-d)} \int_x^{\infty} \frac{dx}{ax^2 +
 (2a + e - b)x + (1-d)}$$

i.e.$$ln (\frac{n_{Sp}}{n_0}) = ln \{ \frac{ax^2}{ax^2 + (2a + e -
b)x + (1-d)}\}^{\frac{1}{2(1-d)}}$$
$$ + \frac{(2a +e -b)d}{(1-d) a \Delta} \ [ tan^{-1} \{ \frac{x +
\frac{(2a + e - b)}{2a} }{\Delta} \} - \frac{\pi}{2}]$$

where $$ \Delta^2 = \frac{1-d}{a} - \{ \frac{(2a + e -b)}{2a} \}^2
> 0$$ otherwise $\Delta$ hence $ln (\frac{n_{Sp}}{n_0})$ will be
imaginary.

Hence,$$ \frac{n_{Sp}}{n_0}=[ \{\frac{ax^2}{ax^2+(2a+e-b)x+(1-d)}
 \}^{\frac{1}{2(1-d)}}]$$
$$ .exp[\frac{(2a+e-b)d}{(1-d)a
\Delta}(tan^{-1}\{\frac{2ax+(2a+e-b)}{2a\Delta} \} -
\frac{\pi}{2})]$$

From equation (3)
$$ dn_E = \{ \frac{(1-a)x^2 + bx + cx + d }{-2ax^2 - x} \} \{ x .
\frac{ (ax^2 - bx +ex - d)}{(1-a)x^2 + (b + c)x +d } dn_E +
\frac{n_{Sp}}{x} dx \}$$

i.e.$$[1 + \frac{ax^2 +(e - b)x - d}{2ax + 1}] dn_E = - \frac{
n_{Sp} \{(1 -a )x^2 +(b + c)x +d \}}{x^2 (2ax + 1)} dx $$

i.e. $$ dn_E = - \int_{\infty}^x \frac{n_{Sp} \{ (1-a) x^2 +(b +
c) x +d \}}{x^2 \{ ax^2 + (2a + e - b)x + 1 - d\}}$$

i.e.$$ dn_E =  \int^{\infty}_x \frac{n_{Sp} \{ (1-a) x^2 +(b + c)
x +d \}}{x^2 \{ ax^2 + (2a + e - b)x + 1 - d\}}$$

Again dividing equation (5) by equation (1) and using $$ x =
\frac{n_{Sp}}{n_{S0}}$$

$$ \frac{dn_{dE}}{dn_{Sp}} = - \frac{ \{k_1a + k_2 (1-a) \} x^2 +
[k_3 b + k_4 c + k_5 \{1-(b + c + e ) \}]x + d k_6 + (1-d) k_7
}{2ax^2 + x}$$

i.e. $$ dn_{dE} = - \frac{(Ax^2 + Bx + C)}{x (2ax + 1)}  \ [ x
\frac{dn_{Sp}}{dn_{dE}} . dn_{dE} + \frac{n_{Sp}}{x} ] $$

where $$ A = k_1^{'} a + k_2^{'} (1 - a) $$
      $$ B = k_3^{'} b + k_4 ^{'} c + k_5 ^{'} \{ 1 - (b + c + e)
      \}$$
      $$ C = dk_6^{'} + (1-d) k_7^{'}$$

      Hence , $$dn_{dE} = - \frac{(Ax^2 + Bx + c )}{x(2ax + 1)} [
      x \frac{(ax^2 - bx + ex - d)}{(Ax^2 + Bx + c )} dn_{dE} +
      \frac{n_{Sp}}{x} dx ]$$

      i.e. $$ \{ \frac{ax^2 + (2a + e - b)x - d + 1 }{(2ax + 1)}\}
      dn_{dE} = - \frac{(Ax^2 + Bx + C ) n_{Sp} }{x^2 (2ax + 1)}
      dx $$

      i.e. $$ n_{dE} = \int _x ^{\infty} \frac{(Ax^2 + Bx + C)
      n_{Sp}}{ x^2 \{ ax^2 + (2a + e - b)x + (1-d)\}}dx$$

      Similarly the integral follows for $n_{dIrr}$. The system of
      equations (6) - (10) have implicit solution because each
      time during evaluation of the integrals (8) - (10) values of
      $n_{Sp}$ are used which is a dependent variable of x.

\clearpage
\begin{figure}
\centering
\includegraphics[width=1\textwidth]{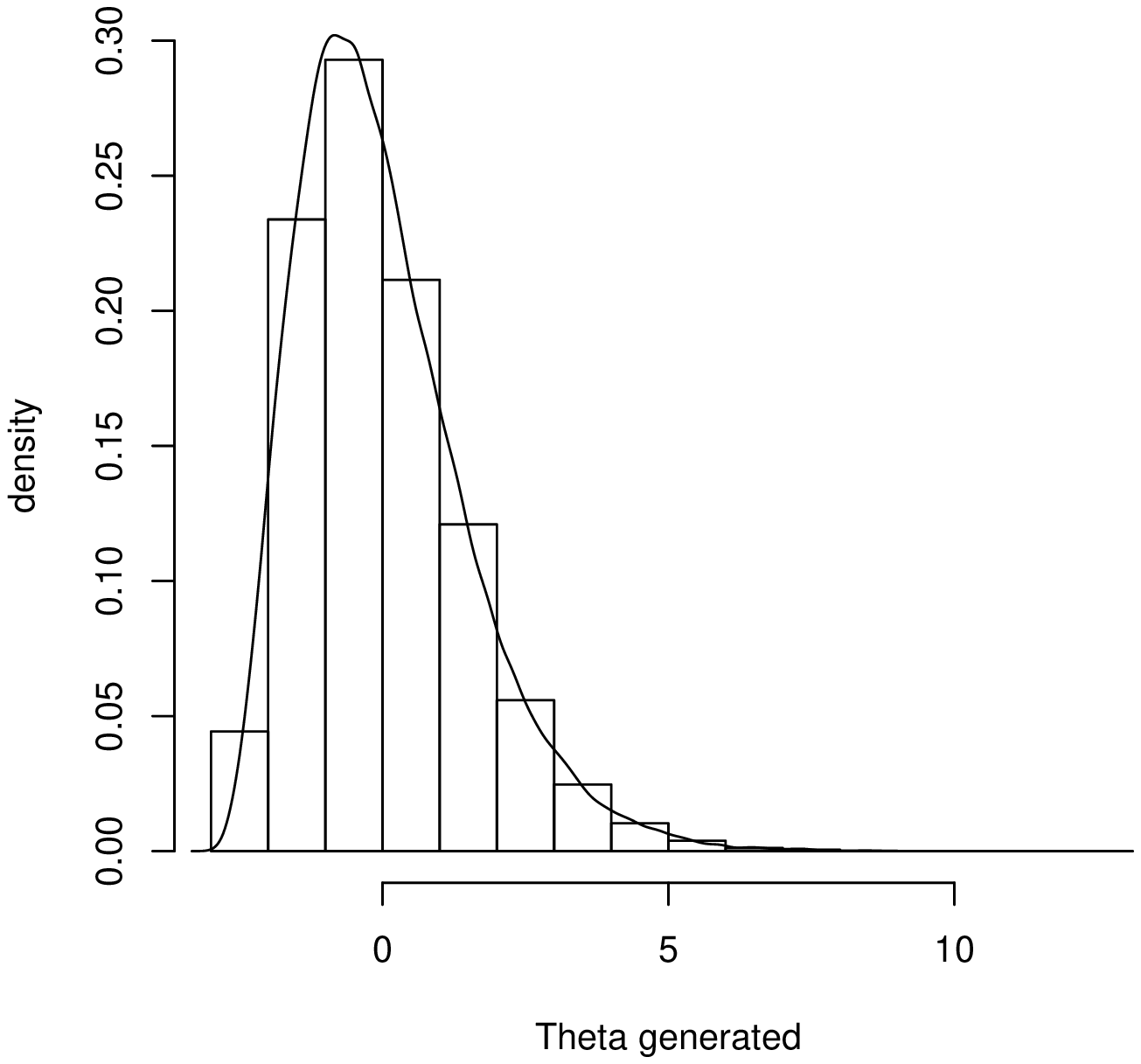}
\caption{ The generated values of $\Theta$ from Gamma
distribution.}
\end{figure}

\clearpage

\begin{figure}
\centering
\includegraphics[width=1\textwidth]{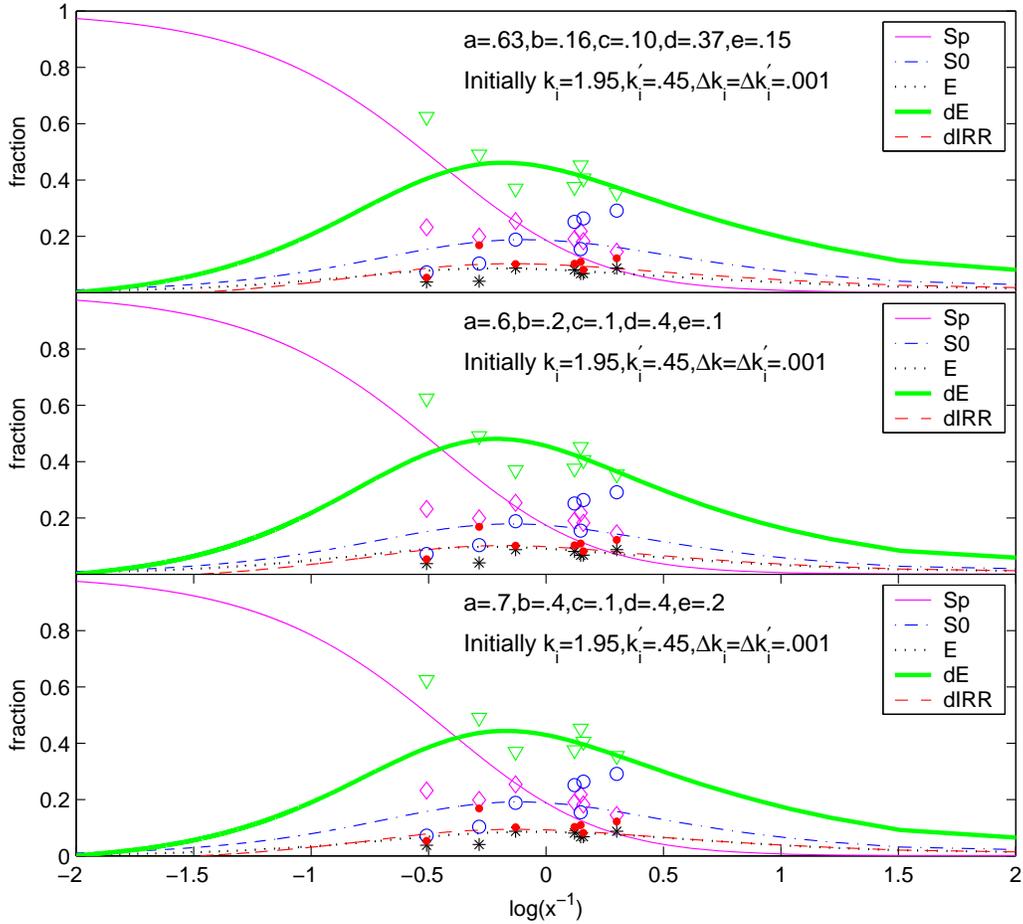}
\caption{ Simulated values of the fractions of galaxies of various
morphological types with observed values for some of the values of
inpact parameters as a function of log(1/x). Magenta open diamonds
are for Spiral galaxies, blue open circles are for S0 galaxies,
black stars are for Ellipticals, green open triangles are for dEs
and red solid circles are for dIrrs.}
\end{figure}

\clearpage

\begin{figure}
\centering
\includegraphics[width=1\textwidth]{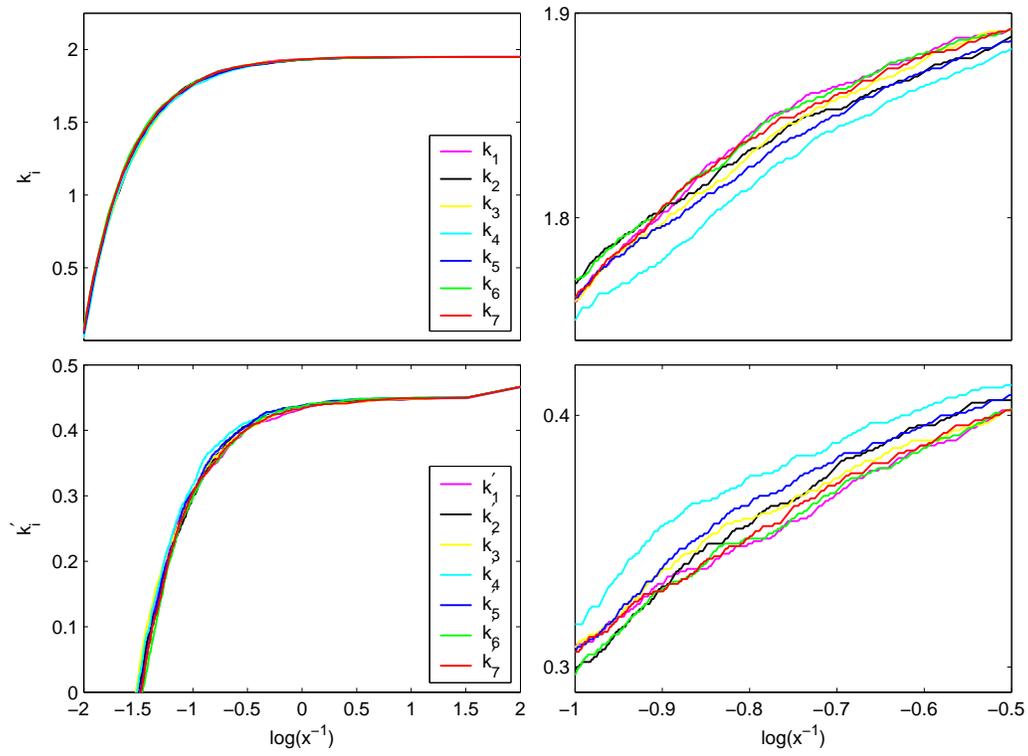}
\caption{Simulated values for $k_i$, the number of dEs formed, and
$k_i^{'}$, the number of dIrrs formed, (i=1-7), as a function of
space density.}
\end{figure}
\clearpage

\begin{table}
\caption{Observed values of $\Theta$ used from Karachentsev et
al.(2004) in Chattopadhyay et al.(2010)}
\begin{tabular}{cccccc}
\hline\hline Galaxy names& $\Theta$ & T&Galaxy names& $\Theta$ &
T\\\hline
E349-031 &   0.5 &10&UGC7242& 0.4 &10 \\
 E410-005 &   0.4& -1&DDO113 & 1.6 &10 \\
 E294-010 &   1.0& -3& DDO125& -0.9& 10 \\
KDG2 &0.4& -1 &UGC7605& 0.7& 10 \\
E540-032 &   0.6 &-3&E381-018 &-0.6& 10 \\
UGC685 & -1.6  &  9& E443-09& -0.9& 10 \\
KKH5  &  -1.2 &10 & KK182 &  1.2 &10 \\
KKH6& -0.8 &10 & UGC8215& -0.5 &10 \\
KK16 &   -0.4 &10 & E269-58 &1.9& 10 \\
KK17 &   -0.3 &   10 & KK189  & 2.0& -3 \\
KKH34& -0.8 &10 & E269-66& 1.7& -1 \\
E121-20 &-1.6 &10 & KK196 &2.2& 10 \\
E489-56& -2.1& 10 & KK197& 3.0 &-3 \\
KKH37 &  -0.3 &10 & KKs55& 3.1& -3 \\
UGC3755& -2.1 &10 & 14247& 1.5& 10 \\
E059-01& -1.5& 9 &UGC8508& -1.0& 10  \\
KK65& -2.0 &10 &E444-78 &2.1& 10 \\
UGC4115& -1.7 &10& UGC8638& -1.3 &10 \\
DDO52&-1.5 &10& KKs57& 1.8 &-3 \\
D564-08 &-1.9  &  10 & KK211& 1.5 &-5 \\
D565-06 &-1.8 &10& KK213 &1.7 &-3 \\
KDG61  & 3.9& -1 & KK217 &1.1 &-3 \\
KKH57& 0.7& -3& CenN &0.9& -3 \\
HS117  & 1.9& 10& KKH86 &-1.5 &10 \\
UGC6541 &-0.7& 10& UGC8833& -1.4& 10 \\
NGC3741 &-0.8& 10 & E384-016& 0.3& 10 \\
E320-14& -1.2& 10 & KK230& -1.0& 10 \\
KK109  & -0.6 &10& DDO190& -1.3& 10 \\
E379-07& -1.3& 10 & E223-09 &-0.8 &10 \\
NGC4163& 0.1& 10 & IC4662 &-0.9& 9\\\hline
\end{tabular}
\end{table}
\clearpage

\begin{table}
\caption{Merger scheme}
\begin{tabular}{cccc}
\hline\hline
 Serial no. &Galaxy type&Probability& Galaxy formed\\
 1 &Sp + Sp $\rightarrow$ &(a) & S0 + $k_1$ dE / $k_1^{'}$ dIrr \\
  &                       &(1-a)& E + $k_2$ dE / $k_2^{'}$ dIrr
  \\\hline
 2& Sp + S0 $\rightarrow$& (b) & E + $k_3$ dE / $k_3^{'}$ dIrr \\
  &                      & (c) & E + S0 + $k_4$ dE / $k_4^{'}$ dIrr\\
  &                      & (e) & S0 + S0 +  $k_8$ dE / $k_8^{'}$ dIrr\\
  &                      & 1 - (b+c+e)& S0 + $k_5$ dE / $k_5^{'}$
  dIrr\\\hline
 3& S0 + S0 $\rightarrow$& (d)& E + $k_6$ dE / $k_6^{'}$ dIrr\\
  &                      & (1-d)& S0 + S0 + $k_7$ dE / $k_7^{'}$ dIrr\\\hline
\end{tabular}
\end{table}

\clearpage
\begin{table}
\caption{Observational numbers of galaxies of various
morphological types in clusters of galaxies.}
\begin{tabular}{ccccccccc}
\hline\hline
 Data set & Cluster& E&
S0& Sp& dE& dIrr& x& log(1/x)\\

1& Virgo&20& 53 & 102 & 252 & 87 & 1.92 & -0.284\\
 & Coma0   & 37 & 115 & 87 & 172 & 47 & 0.7565& 0.12118 \\
 & Coma1& 15 & 50 & 25 & 61& 21& 0.5& 0.30103\\
 & Coma2&10& 39& 27& 60& 12& 0.69& 0.1597\\
 & Coma3& 12& 26& 35& 51&14& 1.3461& -0.12909\\ \hline
2& Centaurus& 16 & 37 & 52& 108& 26& 0.7838& 0.1478\\ \hline
3&VirgoII& 60 & 114& 367& 960&82& 3.21& -0.5078\\\hline

\end{tabular}
\end{table}

\clearpage

\begin{table}
\caption{Best fit Chisquare estimates and corresponding ranges of
parameters}
\begin{tiny}
\begin{tabular}{ccccccccc}
\hline\hline a& b&c&d& e& $k_i$(i=1-7)& $k_i^{'}$(i=1-7)&
$\chi^2$& p-value\\
 0.55 - 0.80& 0.10 - 0.55& 0.05 -
0.70 & 0.35 - 0.40 & 0.05 - 0.40 & Fig. 3 & Fig. 3 &0.9$\pm$0.15&0.995 \\
\hline
\end{tabular}
\end{tiny}
\end{table}

\begin{table}
\caption{Observed number of dwarf galaxies in individual galaxy
mergers}
\begin{tabular}{ccc}
\hline\hline
HII Regions of TDG cadidates& Number& Reference\\

1& 12& Ferreiro et al. 2005\\
2& 11& \\
3& 7& \\
4& 4& \\
5& 4& \\
6& 4 & \\
7& 5& \\
8& 3& \\
9& 3& \\
10& 2& \\
11& 1& \\
Stephan's quintet& 5& Oliverira et al. 2000\\
NGC4922& 3 & Sheen et al. 2009\\
NGC4038/9& 2& Hibberd \& Higdon 2001\\
NGC3227/3226& 1& Mundell et al. 2004\\
NGC7252 & 2 & Schweizer 1982 \\
ESO 148 - IG02 & 3 & Bergvall \& Johansson 1985 \\
The Superantennae NGC4038/4039 & 9 & Mirabel et al.1991\\
Arp 105 & 2 & Duc \& Mirabel  1994 \\
NGC2782 & 1 & Yoshida et al. 1994\\
NGC5291& 11 & Duc \& Mirabel 1998\\\hline

\end{tabular}
\end{table}

\end{document}